\documentclass[
reprint,
superscriptaddress,
amsmath,amssymb,
aps,
prl
]{revtex4-2}

\usepackage{graphicx}% Include figure files
\usepackage{dcolumn}% Align table columns on decimal point
\usepackage{bm}% bold math
\usepackage{hyperref}% add hypertext capabilities
\usepackage[mathlines]{lineno}% Enable numbering of text and display math
\begin{document}

%\preprint{APS/123-QED}

%--------------------------------------------------
% Title and Author list
\title{Diffraction of walking drops by a standing Faraday wave}% Force line breaks with \\

\author{Bauyrzhan K. Primkulov}
\thanks{contributed equally}
\affiliation{Department of Mathematics, MIT}

\author{Davis J. Evans}%
\thanks{contributed equally}
\affiliation{Department of Mathematics, MIT}

\author{Valeri Frumkin}%
\thanks{contributed equally}
\affiliation{Department of Mechanical Engineering, Boston University}

\author{Pedro J. S\'aenz}
\affiliation{Department of Mathematics, UNC, Chapel Hill}

\author{John W.M. Bush}
\email{bush@math.mit.edu}
\affiliation{Department of Mathematics, MIT}

\date{\today}% It is always \today, today,

%******************************************************
% ABSTRACT
\begin{abstract}
The Kapitza-Dirac effect is the diffraction of quantum particles by a standing wave of light. We here report an analogous phenomenon in pilot-wave hydrodynamics, wherein droplets walking across the surface of a vibrating liquid bath are deflected by a standing Faraday wave. We show that, in certain parameter regimes, the statistical distribution of the droplet deflection angles reveals a diffraction pattern reminiscent of that observed in the Kapitza-Dirac effect. Through experiments and simulations, we show that the diffraction pattern results from the complex interactions of the droplets with the standing wave. Our study highlights non-resonant effects associated with the detuning of the droplet bouncing and the bath vibration, which are shown to lead to drop speed variations and droplet sorting according to the droplet's phase of impact. We discuss the similarities and differences between our hydrodynamic system and the discrete and continuum interpretations of the Kapitza-Dirac effect, and introduce the notion of ponderomotive effects in pilot-wave hydrodynamics.
\end{abstract}

\maketitle

%******************************************************
% Introduction to Kapitza-Dirac

%------Walking drops intro-------------
\textbf{Introduction.} 
Diffraction experiments have been pivotal in advancing our understanding of both light and matter. Pioneering light diffraction experiments were conducted in 1665 by Grimaldi, who documented wave properties of light and put forward an early version of the Huygens principle~\citep{grimaldi_physico-mathesis_1665, huygens_traite_1690}. The opposing “corpuscular” view of light pioneered by Newton successfully rationalized light refraction and reflection phenomena~\citep{newton_opticks_1704} and dominated the scientific consensus of the time despite its inconsistency with diffraction experiments. Young’s refinement of diffraction experiments~\citep{young_i_1804} combined with Fresnel’s formal theory that accurately predicted diffraction results~\citep{fresnel_premier_1815}, provided strong support for the wave nature of light. Following Maxwell’s derivation of the relationship between electricity, magnetism, and the speed of light, the particle view was almost entirely abandoned, only to be resurrected again with Einstein’s 1905 explanation of the photoelectric effect based on Planck’s hypothesis of light quanta~\citep{einstein_heuristic_1905}. 
In 1924, de Broglie suggested that wave properties are intrinsic to both light and matter~\citep{de_broglie_recherches_1924}. Specifically, in his theory of the double-solution, he proposed that particles possess internal oscillations that generate a pilot wave that guides their motion. De Broglie thus posited `matter waves': particles have an associated frequency and wavelength. His theory was supported by Davisson and Germer’s 1928 experiments of electron diffraction by a nickel-crystal lattice~\citep{davisson_reflection_1928}, on the basis of which he won the Nobel prize in 1929. 

In 1933, Kapitza and Dirac~\citep{kapitza_reflection_1933} proposed that, just as light may be diffracted by matter, matter may be diffracted by light. They considered a standing wave of light as a diffraction grating for a beam of electrons and predicted that electrons would experience quantized deflection angles due to discrete photon absorption and the subsequent stimulated Compton scattering~\citep{batelaan_colloquium_2007}. The first experimental realization of the Kapitza-Dirac effect was achieved in 1986 in the scattering of sodium atoms by a near-resonant standing-wave laser field~\citep{gould_diffraction_1986}. Since then, the Kapitza-Dirac effect has been observed in a growing number of physical systems~\citep{batelaan_colloquium_2007}, using atoms~\citep{martin_bragg_1988}, electrons~\citep{bucksbaum_high-intensity_1988, lin_ultrafast_2024, freimund_observation_2001, freimund_bragg_2002}, Bose-Einstein condensates~\citep{ovchinnikov_diffraction_1999}, and even complex molecules~\citep{nairz_diffraction_2001}. Today, the effect can be most generally defined as the ``diffraction of a particle by a standing wave''~\citep{batelaan_colloquium_2007}, prompting the question of whether a comparable diffraction effect can also be achieved in a classical system.

A classical realization of de Broglie's matter waves was discovered in 2006 by \citet{couder_single-particle_2006}. \citet{couder_walking_2005} found that a millimetric drop of silicone oil may self-propel on the surface of a vertically vibrated liquid bath via a resonant interaction with its own guiding or `pilot' wave field. The drop and its pilot wave comprise a compound entity, termed a `walker', that represents a macroscopic realization of the physical picture envisioned by de Broglie~\citep{de_broglie_recherches_1924} in which a vibrating particle is guided by its pilot wave~\citep{bush_pilot-wave_2015}. \citet{couder_single-particle_2006} directed these walkers towards a submerged wall with two slits and documented a diffraction-like pattern in the deflection angles of the walker trajectories. These experiments have since been revisited both experimentally and theoretically~\citep{andersen_double-slit_2015, pucci_walking_2018, ellegaard_interaction_2020, pucci_single-particle_2024}. Subsequently, various quantum analogs have been established using the hydrodynamic pilot-wave system, including analogs of orbital quantization~\citep{fort_path-memory_2010, harris_droplets_2014}, quantum corrals~\citep{harris_wavelike_2013, saenz_statistical_2017}, tunneling~\citep{nachbin_tunneling_2017, eddi_unpredictable_2009}, Friedel oscillations~\citep{saenz_hydrodynamic_2020}, superradiance~\citep{frumkin_superradiant_2023}, spin lattices~\citep{saenz_emergent_2021}, Anderson localization~\citep{abraham_anderson_2024}, as well as demonstrations of surreal trajectories~\citep{frumkin_real_2022}, and interaction-free measurement~\citep{frumkin_misinference_2023}.

In this study, we use the walking droplets to establish a hydrodynamic analog of the Kapitza-Dirac effect. We show that as walkers pass through a locally excited standing Faraday wave, the distribution of the deflection angles of their trajectories exhibits statistical diffraction similar to that arising in the Kapitza-Dirac experiments. We demonstrate the emergence of ponderomotive effects in our hydrodynamic system and compare them to those invoked in the interpretation of the Kapitza-Dirac effect.

%-------------------------
% Fig 1
\begin{figure}
    \centering
    \includegraphics[width=.9\linewidth]{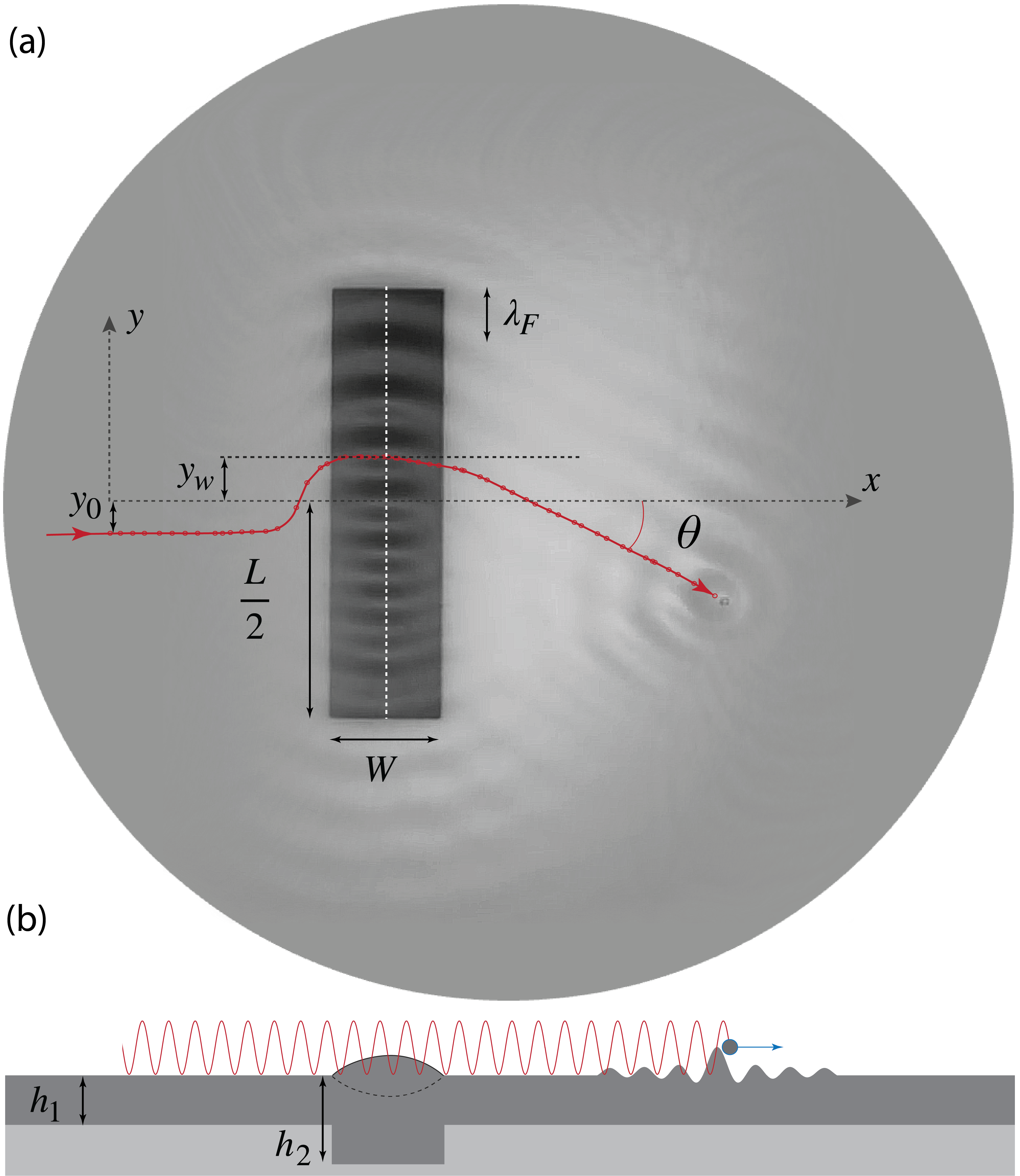}
    \caption{(a)~Top view and (b)~schematic side view of the hydrodynamic system, in which a millimetric walking drop crosses a relatively deep rectangular region ($L=43$\;mm, $W=11$\;mm, $h_1=1.42$\;mm, $h_2=7.92$\;mm) that supports a standing Faraday wave field with wavelength $\lambda_F$. The droplet trajectory is deflected by an angle $\theta$ downstream of the standing wave. The top image is taken with a semi-reflective mirror: bright areas within the rectangular well correspond to the extrema in the wave field, where the liquid surface is flat. The impact parameter and the crossing distance are denoted by $y_0$ and $y_w$, respectively. The latter is the $y$-coordinate of the drop when it crosses the well's centerline (dashed white line). }
    \label{fig:schem}
\end{figure}

%-------------------------
% Fig 2
\begin{figure}
    \centering
    \includegraphics[width=\linewidth]{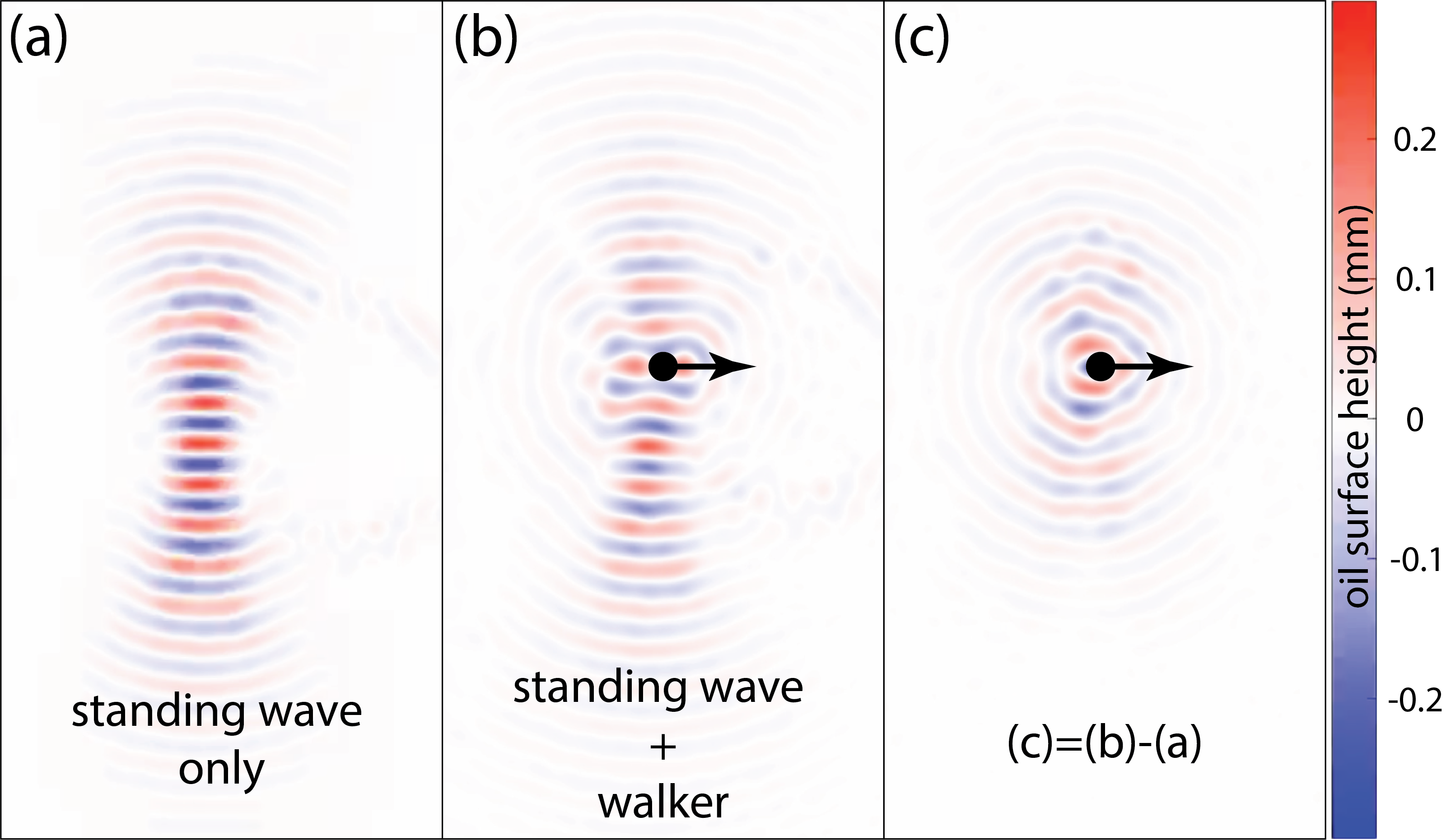}
    \caption{Experimentally-measured surface wave heights acquired directly with a Fast Checkerboard Demodulation method~\citep{wildeman_real-time_2018}. (a)~Standing wave in the absence of the drop. (b)~Standing wave with the drop traversing left-to-right. (c)~The difference in wave heights between images (a) and (b) represents the anomalous pilot wave of the droplet. The black dot marks the droplet position.}
    \label{fig:superpos}
\end{figure}

%******************************************************
% Experimental details
\textbf{Experimental details}. 
We subject a silicone oil bath ($\rho=0.95\;$g/mL, $\nu_o=20\;$cSt, $\sigma=20.6\;$mN/m) to periodic vertical forcing acceleration $\gamma\cos\omega t$ with $f={\omega}/{2\pi}=72$\;Hz (Fig.~\ref{fig:schem}). The bath has depth $h_1=1.42\pm0.05\;$mm and a rectangular well ($L=43$ by $W=11$\;mm) that extends to $h_2=7.92\pm0.05\;$mm depth. The dimensions of the well are adjusted to produce a standing wave with crests aligned only in a direction perpendicular to the long edge of the well. (As a point of comparison, a large square well would instead generate a checkerboard Faraday wave pattern). We force the bath at a peak acceleration ($\gamma=3.74g$) that excedes the Faraday threshold immediately above the rectangular well (for which $\gamma_F=3.72g$), so that a stable sub-harmonic standing wave ($\omega_F={\omega}/{2}=36$\;Hz) forms only in the vicinity of the well (see videos 1 and 2). As is evident in Fig.~\ref{fig:schem}a, the standing wave extends substantially (2-3 $\lambda_F$) beyond the well and has a wavelength of $\lambda_F=5.16\;$mm. Its characteristic amplitude is $h\approx0.24$\;mm. The wave has a relatively narrow waist (Fig.~\ref{fig:superpos}a) and its shape resembles that of the laser beam in the diffraction regime of the Kapitza-Dirac experiment~\citep{batelaan_colloquium_2007}. Although the expected Faraday wavelengths for the shallow and deep regions differ only slightly (5.04 and 5.16 mm, respectively)~\citep{benjamin_stability_1954}, the measured $\lambda_F$ matches that of the deeper region. The Faraday threshold for the shallow region of the bath was measured to be $3.80g$.

We isolated our system from ambient air currents with a transparent lid~\citep{harris_droplets_2014}, and used a launcher to direct $0.82\pm0.02$\;mm-diameter drops with a free-walking speed of $14.7\pm1.0\;$mm/s and prescribed impact parameter $y_0$ (i.e. vertical offset from the well center) towards the standing wave (see Fig.~\ref{fig:schem}a). We recorded the drop's horizontal trajectories using a CCD camera mounted above the fluid bath. We alternated between three imaging settings to capture various aspects of our experiments. In the first setting, the color of the bath's base was set to black beneath the rectangular well and white elsewhere. This was done to make clear the extent of the rectangular well. Additionally, a semi-reflective mirror was placed at a 45-degree angle in front of the camera in order to image the form of the surface waves (see Fig.~\ref{fig:schem}a). In the second setting, we removed the semi-reflective mirror and set the entirety of the bath's base to black. Here, the drop appears as the only bright spot, allowing for optimal tracking of its trajectory (see video 3). In the third setting, we used a black-and-white checkerboard pattern on the bath's base in order to make quantitative measurements of the liquid surface wave height through the Fast Checkerboard Demodulation (FCD) method~\citep{wildeman_real-time_2018}.

\begin{figure*}
    \centering
    \includegraphics[width=\linewidth]{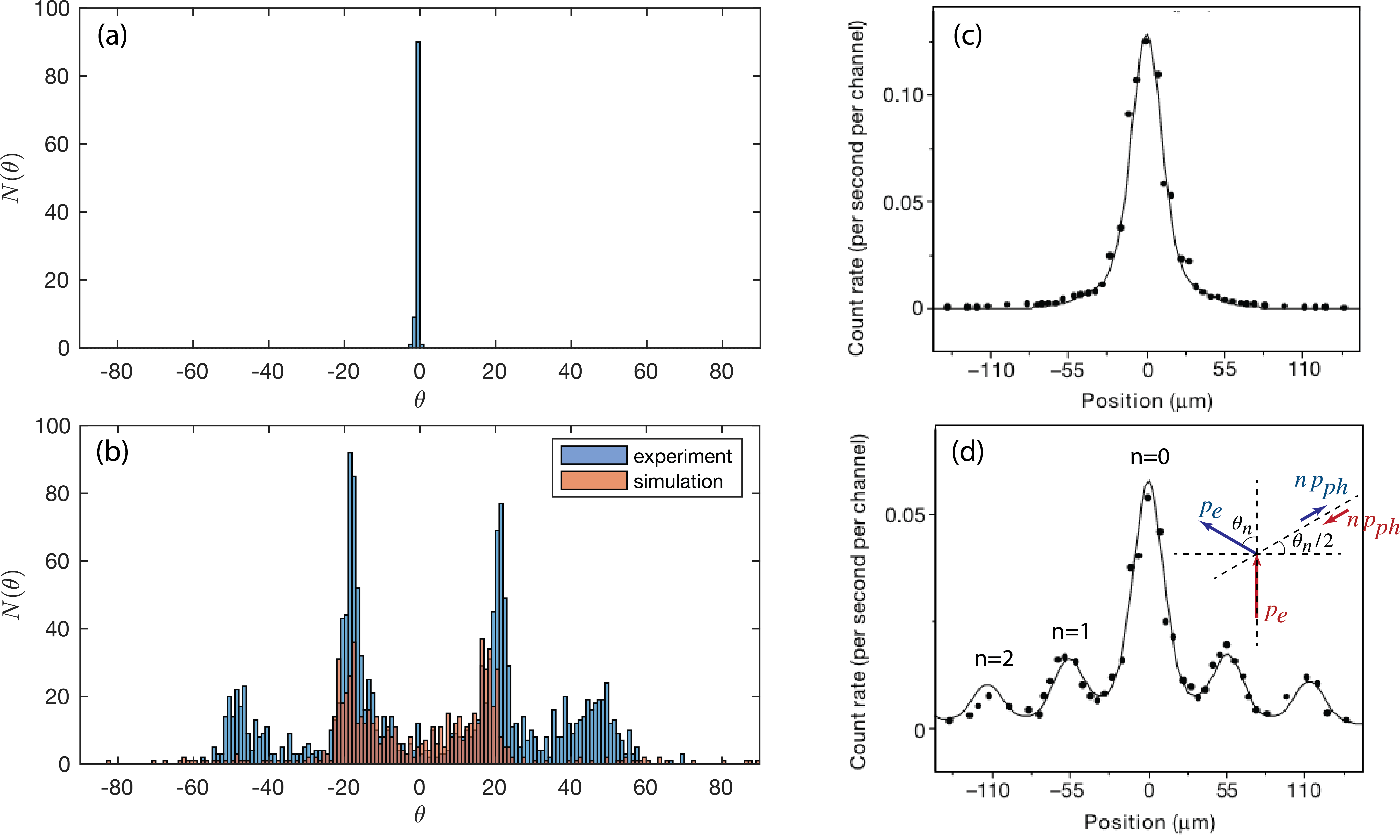}
    \caption{Histogram of the deflection angles (a) without and (b) with the standing wave above the rectangular well. Experiments in (a) were conducted below the Faraday threshold, where drop trajectories were unaffected by the well. The deflection angles $\theta$ in (b) were deduced from 1623 experiments (blue) and 690 simulations (red), for which the impact parameter was $-5\leq y_0\leq5$\;mm. Simulations capture two central peaks but not the outer two owing to simplifications in the wave model. Plots (c) and (d) show an electron-deflection histogram with counter-propagating laser beams off and on, respectively (reprinted from the KD experiments of \citet{freimund_observation_2001}). The electron beam deflection order $n$ corresponds to the number of photon recoil events~\citep{kapitza_reflection_1933, batelaan_colloquium_2007}, during each of which the electron has a momentum $p_{ph}$ transferred to it by the standing wave (see inset). The central peaks evident in both (c) and (d) indicate that a significant portion of electrons remains unaffected by the momentum exchange events. Conversely, in our hydrodynamic experiments, most trajectories are substantially deflected by the standing wave.}
    \label{fig:angle_hist_exp}
\end{figure*}

\textbf{Experimental observations}. 
% Experiments histogram - sweep of yp
The Kapitza-Dirac effect with electrons, as demonstrated by \citet{freimund_observation_2001}, is typically observed by comparing the electron distribution at the detector in the presence and absence of the standing wave field. We replicate this approach in our hydrodynamic system and present our findings alongside the KD experiment results from \citet{freimund_observation_2001} in Fig.~\ref{fig:angle_hist_exp}. In our experiments, we launched individual walkers, 1623 in total, towards the standing Faraday wave, and measured their deflection angle $\theta$ from the horizontal. Fig.~\ref{fig:angle_hist_exp}a shows the statistical distribution obtained in our experiments by sweeping a range of walker impact parameters in the range $-5\leq y_0\leq5$\;mm (see Appendix, Fig.~\ref{fig:impact_par}). The resulting statistics reveals a diffraction-like pattern with four clear peaks, resembling the electron deflection statistics in the Kapitza-Dirac experiment~\citep{freimund_observation_2001} (see Fig.~\ref{fig:angle_hist_exp}c).  

\begin{figure}
    \centering
    \includegraphics[width=\linewidth]{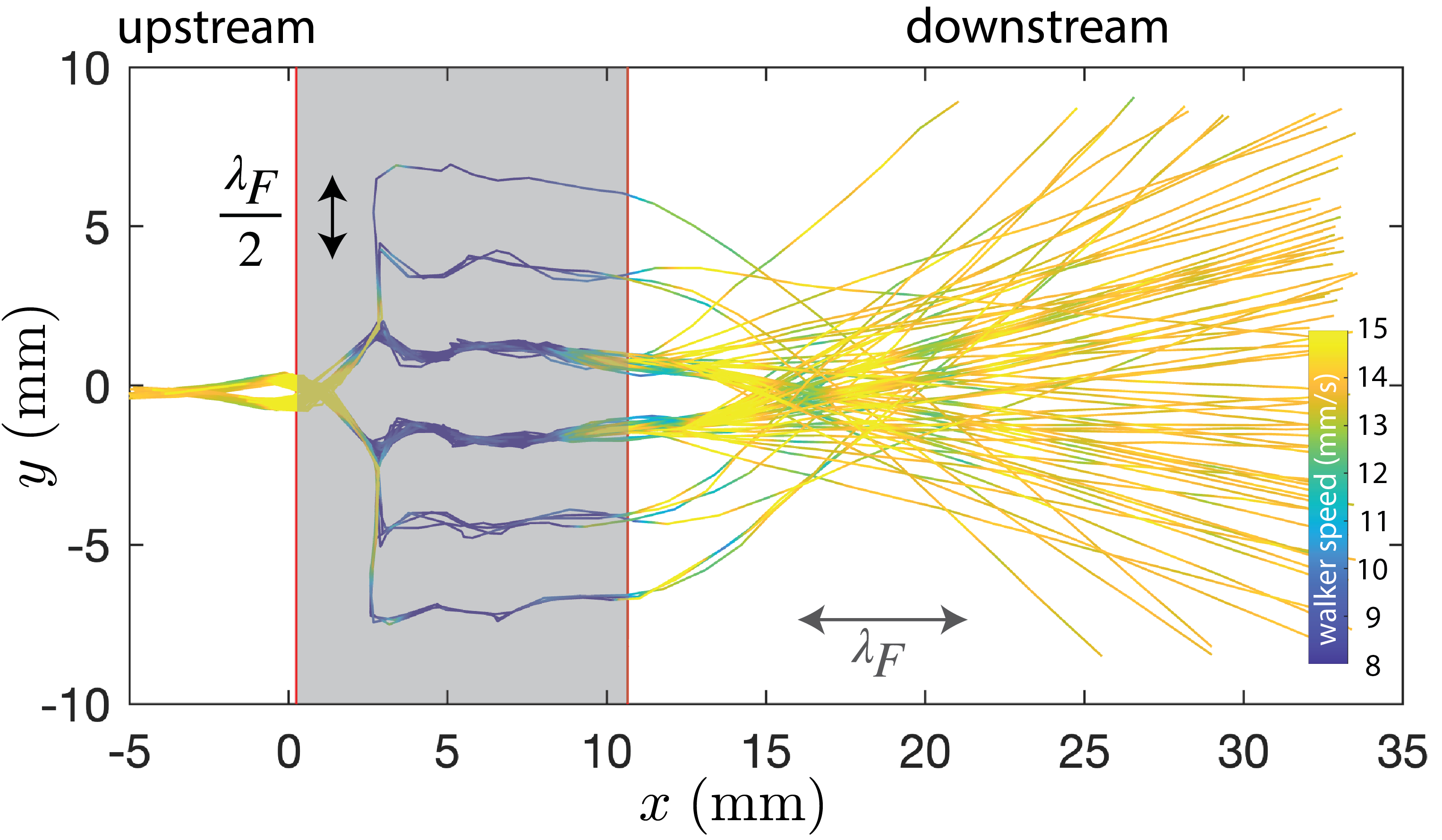}
    \caption{An ensemble of experimental droplet trajectories with $y_0 = -0.21\pm 0.08$\;mm. The initial ensemble splits into one of six distinct trajectories, each of which follows an extrema in the well's standing wave field. Trajectories are color-coded according to speed. Speed oscillations with wavelength $\lambda_F$ are evident downstream of the well.}
    \label{fig:tracks}
\end{figure}

In contrast to the quantum Kapitza-Dirac experiment, where only the final position of the diffracted particles is registered at the detector, our system allows us to directly observe the trajectories of the diffracted particles, thus providing additional insight into the system dynamics. For example, we can assess the sensitivity of walker trajectories to their initial conditions by fixing the impact parameter $y_0$ to a single value. Fig.~\ref{fig:tracks} shows an ensemble of experimental walker trajectories with impact parameter $y_0=-0.21\pm 0.08$\;mm, which is very close to the midpoint of the rectangular well. Although these walkers have essentially the same $y_0$, the walker ensemble splits into six distinct tracks that are separated by $\lambda_F/2$ and overlay the extrema of the standing wave. These trajectories subsequently exhibit a wide range of deflection angles in the range $-50^\circ \lesssim \theta \lesssim 50^\circ$, demonstrating the complex nature of walker-standing-wave interactions.

%-------------------------
% Fig 3

\begin{figure*}
    \centering
    \includegraphics[width=\linewidth]{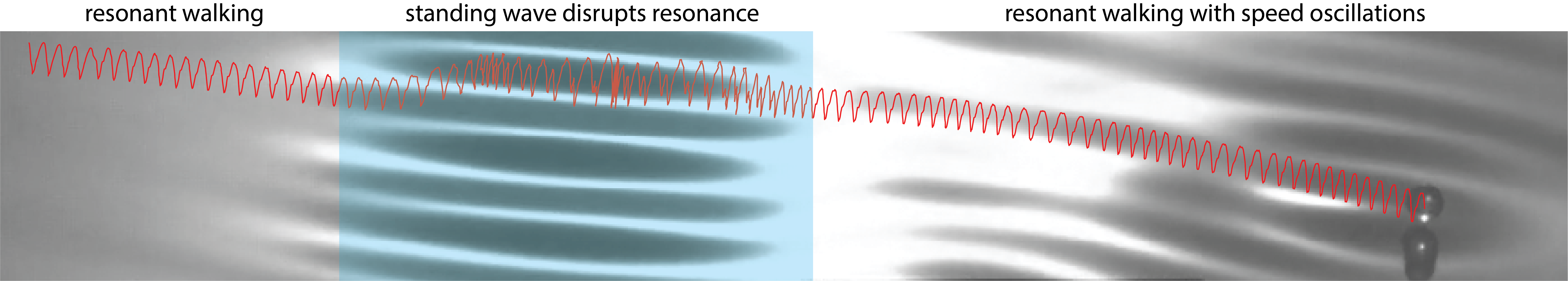}
    \caption{A standing wave disrupts the resonance of the walking drop, specifically the periodicity of its vertical dynamics. This disruption is revealed by the high-speed imaging of the trajectory of the droplet's center of mass (red line), and is accompanied by the horizontal speed oscillations evidenced by the variation in step size per bounce above and downstream of the well. See Fig.~\ref{fig:highspeed_sim} in the Appendix for the comparison with the simulation.}
    \label{fig:highspeed}
\end{figure*}

Our experiments also reveal that walkers exhibit significant speed modulations as they pass through the standing wave. First, the speed of walkers above the standing wave is reduced by approximately $50\%$ relative to that outside the well. High-speed images of the walker trajectories from an oblique angle show that the drop's bounces are periodic upstream of the well but become erratic when the drop encounters the standing wave (Fig.~\ref{fig:highspeed} and video 4). This disruption of the drop's periodicity is responsible for the anomalous slowing of the walkers above the standing wave. Second, Fig.~\ref{fig:tracks} reveals that walkers undergo speed oscillations after passing the well. These underdamped speed oscillations~\citep{durey_speed_2020} produce a roughly sinusoidal variation of speed as a function of $x$, which will lead to a statistical signature comparable to that arising in the hydrodynamic analog of Friedel oscillations~\citep{saenz_hydrodynamic_2020}.

%-------------------------
% Fig 4

%-------------------------
% Fig 5

%******************************************************
% Model description
\textbf{Model}. Much of the rich pilot-wave dynamics observed in our experiments can be rationalized through the non-resonant walker model based on the work of \citet{molacek_drops_2013a, molacek_drops_2013b} and detailed in \citet{primkulov_non-resonant_2024}. The model accurately captures the vertical dynamics and relaxes the commonly imposed condition of resonance between the drop and the bath~\citep{turton_review_2018}, the absence of which is a key feature of our system. After non-dimensionalizing distances by the drop radius $R$ and time by the drop's natural frequency $\omega_D=\sqrt{{\sigma}/{\rho R^3}}$, the equations that govern the walker's vertical and horizontal dynamics may be written as

\begin{align}
    \Ddot{z}_p &= F_N(\tau)-\mathit{Bo},
    \label{eqn:zdot} \\
     \mathbf{\ddot{x}}_p +(\mathcal{D}_h F_N(\tau) + \frac{9}{2}\mathit{Oh}_a) \mathbf{\dot{x}}_p &= - F_N(\tau) \nabla{(h+H)},
    \label{eqn:xdot}
\end{align}

\noindent where $\mathbf{x}_p=(x_p,y_p)$ and $z_p$ are the horizontal and vertical coordinates of the drop's base in the lab frame of reference, respectively~\citep{molacek_drops_2013a}, $\mathit{Oh}_a=\mu_a/\sqrt{\sigma\rho R}$ is the Ohnesorge number (based on the viscosity of air), $\mathit{Bo}={\rho g R^2}/{\sigma}$ is the Bond number. $\mathcal{D}_h=0.17$ is the damping coefficient that accounts for the loss of horizontal momentum during impact~\citep{molacek_drops_2013b}. The normal force $F_N(\tau)$ is non-zero only during impact, and is comprised of spring force and damping force components. The former captures the influence of surface tension, while the latter incorporates that of viscous dissipation. We thus write

\begin{equation}
    F_N(\tau) = -\mathcal{H}(-Z(\tau)) (\mathcal{D}_v \dot{Z}(\tau)+\mathcal{C}_v Z(\tau)),
    \label{eqn:force}
\end{equation}

\noindent where $\mathcal{H}$ is the Heaviside step function. The vertical spring ($\mathcal{C}_v=0.59$) and damping ($\mathcal{D}_v=0.48$) coefficients for our system have been directly measured in previous studies~\citep{molacek_drops_2013a, couchman_bouncing_2019}. The height of the drop base above the liquid bath surface, $Z(\tau)$, may be written as

\begin{equation}
    Z(\tau) = z_p(\tau)-z_b(\tau)-h(\mathbf{x}_p,\tau)-H(\mathbf{x}_p,\tau),
\end{equation}

\noindent where $z_p(\tau)$ is the drop height in a stationary lab frame, $z_b(\tau)$ is the position of the unperturbed bath free surface, $h(\mathbf{x},\tau)$ is the walker's pilot-wave field, and $H(\mathbf{x},\tau)$ is the standing wave above the rectangular well. We use the reduced wave model of \citet{molacek_drops_2013b} for the wavefield $h(\mathbf{x_p},\tau)=\cos{(\Omega \tau/2)} \sum_{i=1}^{n} A_i \text{exp}(-\frac{\tau-\tau_i}{\tau_F \mathit{Me}}) (\tau-\tau_i)^{-1/2} J_0 (k_F |\mathbf{x_p}-\mathbf{x_i}|)$, where $\mathbf{x_i}$ marks the locations of the drop's impacts and $A_i$ prescribes the resulting wave amplitudes (see \citep{primkulov_non-resonant_2024}). Both $h(\mathbf{x},\tau)$ and $H(\mathbf{x},\tau)=\phi(\mathbf{x}) \cos \Omega\tau/2$ are sub-harmonic, so oscillate at the Faraday frequency $\omega_F=\omega/2$~\citep{benjamin_stability_1954,molacek_drops_2013b}.

%*****************************************************
% Note about superposition
We obtain the wave envelope for $H(\mathbf{x},\tau)$ directly from experiments (see Fig.~\ref{fig:superpos}a). Furthermore, we check whether the linear superposition of the standing wave with walker's pilot wave in Eq.~\eqref{eqn:xdot} is a reasonable assumption. Subtracting a standing wave envelope (Fig.~\ref{fig:superpos}a) from a wave field measured with the drop crossing the well (Fig.~\ref{fig:superpos}b) leaves the anomalous pilot wave generated by the drop (Fig.~\ref{fig:superpos}c). The shape of the walker pilot wave field generated above the well (Fig.~\ref{fig:superpos}c) closely resembles that of a free walker~\citep{oza_trajectory_2013}. This correspondence provides some justification for our using linear superposition of $h$ and $H$ in Eq.~\eqref{eqn:xdot}.

\begin{figure}
    \centering
    \includegraphics[width=\linewidth]{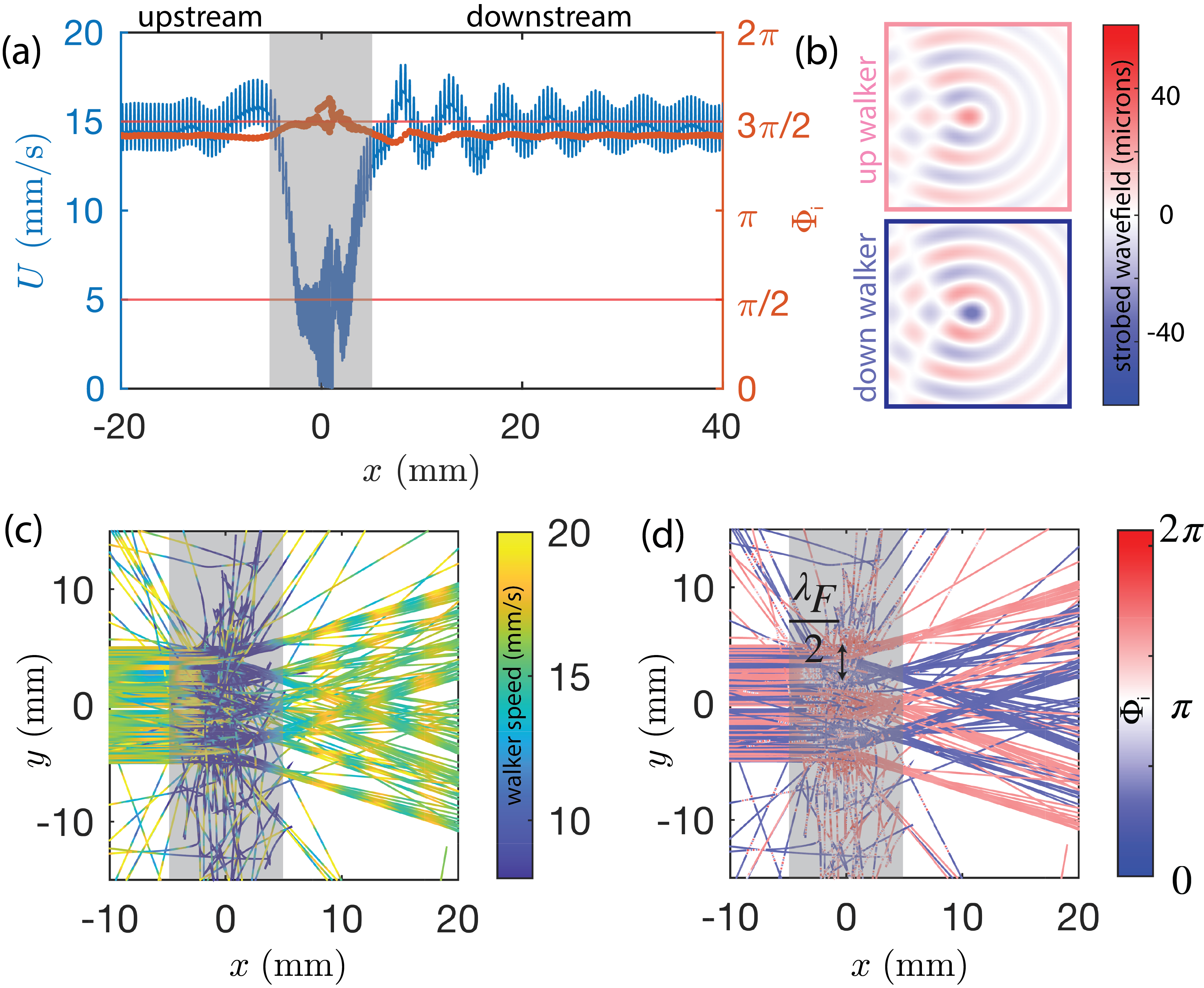}
    \caption{Simulations of a walker crossing the standing Faraday wave field depicted in Fig.~\ref{fig:superpos}a. (a)~Walker speed (blue curve) and the phase of drop-bath impact [eq.~\eqref{eqn:phi}] in relation to the period of the standing wave (orange dots). The two horizontal red lines correspond to the phases at which (i) the fluid bath is moving upwards with the highest speed and (ii) the amplitudes of the pilot and standing waves are zero. The lateral extent of the well is indicated by grey shading. (b)~The drop can assume one of two states based on its phase of impact $\Phi_i$ [eq.~\eqref{eqn:phi}]: either $\Phi_i$ is above (`up') or below (`down') $\pi$. Here the wavefields are strobed at the impact of the `up' walkers. (c)~An ensemble of simulated droplet trajectories colored according to speed indicating in-line oscillations downstream of the well comparable to those reported experimentally in Fig.~\ref{fig:tracks}. (d)~Color coding according to phase $\Phi_i$ indicates dynamic sorting of walkers along different channels according to their vertical bouncing phase.}
    \label{fig:simulations}
\end{figure}

%******************************************************
% Results and discussion
\textbf{Simulations}. 
% Simulations U and Phi
Fig~\ref{fig:simulations}a shows a simulation of a walker traversing the standing wave field. One can keep track of the walker's impact phase relative to the oscillations of the Faraday wave through 

\begin{equation}
    \Phi_i = \frac{\int_{\tau_c} F_N(s)\frac{\Omega s}{2}ds}{\int_{\tau_c} F_N(s)ds}, \hspace{2mm} (\text{mod} \ 2\pi)
    \label{eqn:phi}
\end{equation}

\noindent where $\tau_c$ is the contact time, specifically, the duration of impact between droplet and bath, during which the intervening lubrication layer mediates forces between the two. The impact phases $\Phi_i=\pi/2$ and $\Phi_i=3\pi/2$ correspond to phases at which the upswing of the bath has a maximum speed, and the amplitudes of the pilot $h(\mathbf{x})$ and standing $H(\mathbf{x})$ waves are zero. These specific impact phases, at which one expects the droplet speeds to be minimal, are indicated by two horizontal red lines in Fig.~\ref{fig:simulations}a. 

Walkers maintain constant impact phase $\Phi_i$ when they are in resonance with the vibrational forcing of the bath, for example, in the free-walking state. This constancy is evident in both experiments (Figs.~\ref{fig:highspeed}, \ref{fig:tracks}) and simulations (Fig.~\ref{fig:simulations}a) whenever the drop walks over a quiescent portion of the fluid bath. The model captures a key feature of the experiment; specifically, the walker slows as it passes over the standing wave (see Figs.~\ref{fig:highspeed}, \ref{fig:tracks}).  Here, the walker's speed decreasing with depth results from the system being above $\gamma_F$ in the deep region, and can be readily rationalized through consideration of the drop's impact phase. As evident from the red dots in Fig.~\ref{fig:simulations}a, above the standing wave, the impact phase $\Phi_i$ becomes irregular and approaches the zero-amplitude pilot-wave state $3\pi/2$, which induces a significant speed reduction. 

% Speed oscillations in simulations
Our simulations recover the experimentally-observed speed oscillations with wavelength $\lambda_F$ downstream of the standing wave (Fig.~\ref{fig:tracks}), a feature of pilot-wave hydrodynamics known to result in the emergence of quantum-like statistics~\citep{saenz_statistical_2017, durey_classical_2021}.
% Simulation tracks
Here, the walkers undergo underdamped speed oscillations following the abrupt change prompted by the standing wave field. These speed oscillations have a wavelength of $\lambda_F$ (see Fig.~\ref{fig:simulations}c) and the associated correlation between position and speed results in a commensurate statistical signature in particle position.

% Standing wave sorts up/down walkers
Our model highlights that there are two kinds of walkers (up/down walkers) with identical free walking speeds but with impact phase $\Phi_i$ differing by $\pi$, a distinction that becomes important owing to the droplet's interaction with the subharmonic standing wave (Fig.~\ref{fig:simulations}b)~\citep{primkulov_non-resonant_2024}. A walker starting from rest intermittently switches between the two states until it locks into one of them, either $\Phi_i<\pi$ (down) or $\Phi_i>\pi$ (up). Therefore, essentially half of the walkers approaching the standing wave are in the up state, while the other half is in the down state. At the drop-bath impact, the two kinds of walkers interact with the standing wave at phases differing by $\pi$. Thus, what appears as the maxima of the standing wave for the up walker is a minima for the down walker. In fact, color-coding the walker trajectories with $\Phi_i$ in Fig.~\ref{fig:simulations}d reveals that the standing wave serves to sort up and down walkers, with their respective tracks being separated by $\lambda_F/2$. This is also the case in experiments. The supplemental videos 5 and 6 show the FCD imaging of the wave height during the KD experiments, where up/down drops are sorted into separate tracks separated by $\lambda_F/2$. Notably, all walkers, `up' or `down,' are channelled along minima in the standing wave field.

% Simulations histogram
Simulations corresponding to our experimental conditions recover two central peaks of the deflection angle histogram in Fig.~\ref{fig:angle_hist_exp}. These peaks are essentially dictated by the geometry of the standing wave---one can tune the placement of the peaks by adjusting the width of the standing wave envelope. However, the fact that the numerical results do not recover the two smaller peaks in Fig.~\ref{fig:angle_hist_exp} with the experimentally measured standing wave, suggests that our model does not capture some features of the experiment. Indeed, a comparison of Fig.~\ref{fig:tracks} and Fig.~\ref{fig:simulations}c,d reveals one such discrepancy between simulation and experiments. Specifically, the simulated walker trajectories are not as efficiently channelled along the extrema of the standing wave. Furthermore, supplemental video 7 shows that walkers excite an additional wave mode in the rectangular well, which is tilted relative to the standing wave shown in Fig~\ref{fig:superpos}a. These modes are not included in our theoretical model and may be responsible for the smaller secondary peaks evident in Fig.~\ref{fig:angle_hist_exp}.

%******************************************************
% Conclusion
\textbf{Discussion and Conclusions}. 
The original Kapitza-Dirac diffraction pattern of electrons is commonly interpreted through either the particle-particle interaction picture of Kapitza and Dirac~\citep{kapitza_reflection_1933} (\emph{discrete model}) or the solution of the Schrödinger equation with a ponderomotive potential~\citep{batelaan_colloquium_2007} arising from the electromagnetic standing wave (\emph{continuum model}). While the diffraction pattern in our walker experiments has a purely classical origin and so requires no interpretation, we can examine the similarities and differences between our hydrodynamic system and the two interpretations of the Kapitza-Dirac effect. 

In the \emph{discrete} interpretation of Kapitza and Dirac, the order $n$ of the electron beam deflection angle denotes multiples of photon absorption and subsequent recoil events. These events leave the magnitudes of both the energy and momentum of the photons and electrons unchanged~\citep{kapitza_reflection_1933, batelaan_colloquium_2007}. Consequently, $\sin(\theta_n/2)={n p_\text{ph}}/{p_e}$, where $p_\text{ph}$ and $p_e$ are the incident photon and electron momenta, respectively (see Fig.~\ref{fig:angle_hist_exp}d inset). Therefore, in the experiments of \citet{batelaan_kapitza-dirac_2000} reprinted in Fig.~\ref{fig:angle_hist_exp}, the first two orders of the deflection angles ($n=1, 2$ in Fig.~\ref{fig:angle_hist_exp}c) are below one degree. The momentum exchange events are relatively infrequent, with most electrons experiencing no deflection and only a small fraction undergoing a single-order deflection. In contrast, the deflection angles are considerably higher in our hydrodynamic experiments (Fig.~\ref{fig:angle_hist_exp}). Moreover, the droplet undergoes approximately 50 impacts (or momentum exchange events) as it crosses a standing wave. The resulting deflection angle is determined by the lateral momentum transferred during the entire sequence of impacts with the standing wave crests. 

\begin{figure}
    \centering
    \includegraphics[width=\linewidth]{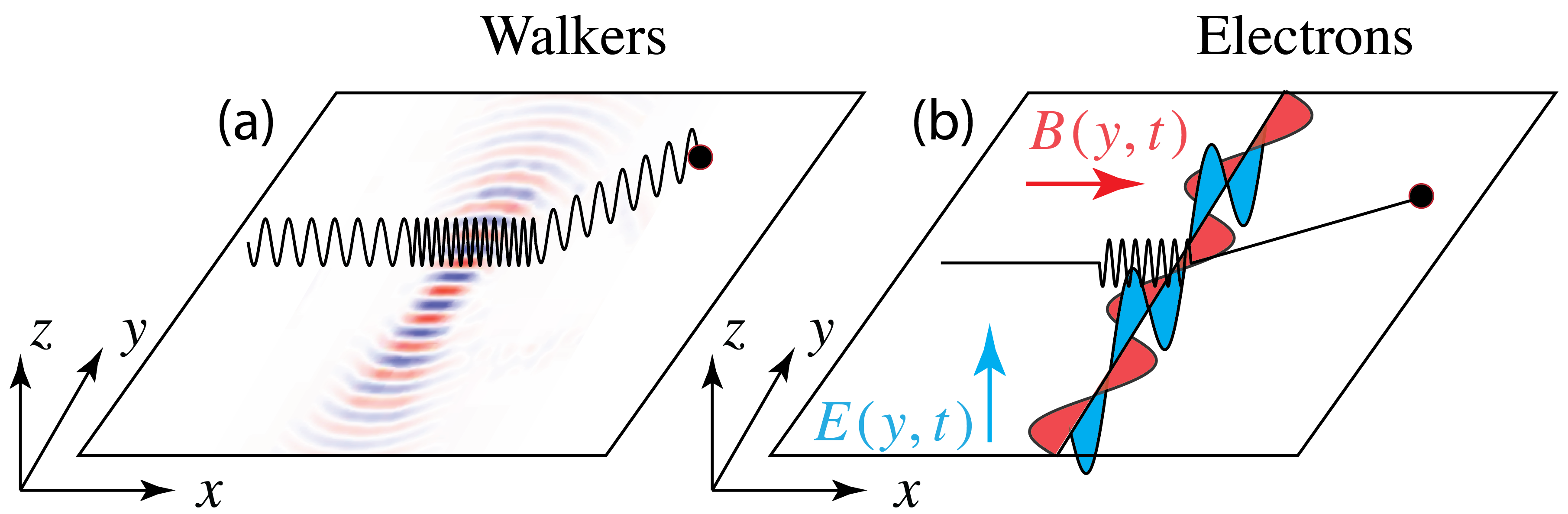}
    \caption{Schematic comparison of ponderomotive effects arising in (a)~walkers moving over a Faraday wave field $H=H(x,y,t)$ and (b)~electrons moving through an electromagnetic standing wave, with electric field $\vec{E}=E(y,t)\hat{k}$ and magnetic field $\vec{B}=B(y,t)\hat{i}$.}
    \label{fig:KDcomparison_scheme}
\end{figure}

Our hydrodynamic system is much closer to the \emph{continuum} interpretation of the Kapitza-Dirac effect described in \citet{batelaan_kapitza-dirac_2000}. To enable comparison between the two, we align the electromagnetic standing wave with the Faraday wave as depicted in Fig.~\ref{fig:KDcomparison_scheme}. An electron moving through this standing wave experiences rapid oscillations along the $z$-axis due to the force from the electric field $E_z = E_0(y)\cos(\omega t)$. Owing to the presence of the magnetic field $B_x = B_0(y)\sin(\omega t)$, the velocity $v_z = \frac{e}{m\omega}E_0(y)\sin(\omega t)$ induces a Lorenz force $F_y = -e v_z B_x$ in the $y$-direction. Despite the cyclic oscillations of $v_z$ and $B_x$ averaging to zero, the Lorenz force $F_y$ does not, owing to the correlations between $v_z$ and $B_x$. The time-averaged force may thus be derived from the ponderomotive potential $U_p = \frac{e^2}{4m\omega^2} E_0(y)^2$~\citep{batelaan_kapitza-dirac_2000}, which drives electrons towards the zeros of the electric field. \citet{freimund_observation_2001} incorporated this ponderomotive potential into the Schrödinger equation, and with a small correction noted in their work, successfully reproduced the diffraction pattern shown in Fig.~\ref{fig:angle_hist_exp}c. 

In the hydrodynamic system, a similar situation arises when averaging the horizontal force during a resonance disruption event. The droplet-bath normal force $F_N(t)$ induces a horizontal force $F_N(t) \cos(\omega_F t) \nabla \phi(\mathbf{x})$ on the droplet due to the presence of the standing wave of height $\phi(\mathbf{x})\cos(\omega_F t)$. When resonance is disrupted, one expects the average of $F_N(t) \cos(\omega_F t)$ to be zero because the impact times are effectively random. As in the KD effect, the net force does not average to zero and can be derived from the ponderomotive potential $U_p = \frac{1}{2}K|\nabla \phi|^2$ (see Appendix). The corresponding vertical and horizontal forces, along with the resulting ponderomotive potentials for both the hydrodynamic and quantum systems, are listed in Table~\ref{tab:tableanalogies}.

\begin{table}
\caption{\label{tab:tableanalogies}%
Comparison between the KD effects arising in the electromagnetic and hydrodynamic systems. We direct the electromagnetic standing wave field along the y-axis, to match the overall direction of the Faraday wave, as in Fig.~\ref{fig:KDcomparison_scheme}. The first and second rows of the table represent forces responsible for vertical oscillations and horizontal drift. The third row shows the ponderomotive potentials responsible for the horizontal deflections arising in the two systems.
}
\begin{ruledtabular}
\begin{tabular}{lcc}
\textrm{}&
\textrm{\textbf{electron}}&\textrm{\textbf{droplet}}\\
\colrule
\textbf{vertical force}&$e E_0(y)\sin(\omega t)$ & $F_N(t) - mg$\\
\hline
\textbf{lateral force} & $ev_z(y,t)B_y(y,t)$ & $F_N(t)\cos(\omega_F t) \nabla \phi$\\
\hline
\textbf{ponderomotive} &$\frac{e^2}{4m\omega^2}E_0(y)^2$ & $\frac{1}{2}K|\nabla \phi|^2$ \\
\textbf{potential}&{}&{}\\
\end{tabular}
\end{ruledtabular}
\end{table}

We have explored the dynamics of a walking drop crossing a standing Faraday wave field. The drop speed is reduced above the wave field and exhibits underdamped oscillations downstream of it. The distribution of the walker deflection angles induced by the standing wave is reminiscent of the diffraction pattern obtained in the Kapitza-Dirac experiment. Our model and experiments reveal that the standing wave serves to sort the walkers according to their impact phase $\Phi_i$. Just as the Stern-Gerlach apparatus sorts atoms according to their up/down spin states, the standing wave in our experiment sorts up/down walkers and sends them in different directions. Finally, we have shown how non-resonant effects may lead to a ponderomotive potential similar in form to that utilized in the statistical modelling of the KD effect with electrons. Consideration of such ponderomotive effects in other hydrodynamic quantum analogs will be the subject of future work.

\bibliography{references}% Produces the bibliography via BibTeX.

\section{Appendix}
\subsection{Idealized Ponderomotive Force}

The dimensional form of the trajectory equation \eqref{eqn:xdot} in the presence of a standing wave (with amplitude $\phi \gg h$) and for short times $m\ddot x \gg \zeta \dot x$ reduces to
\begin{equation}
    m \ddot x = -f(t) \nabla \phi(x),
\end{equation}
where $f(t) = F_N(t)\cos(\omega_F t)$, and $\zeta$ is the damping coefficient. A resonance disruption event (see Fig.~\ref{fig:highspeed}) is a sequence of 5 to 10 bounces ($< 0.2$s) where the impact times are random. We idealize this by assuming $\overline{f} = 0$ when averaged over this time interval. To estimate the net (average) force $\overline{f(t)\nabla \phi(x)}$ over the time interval, we approximate the wave by linearizing about a stationary point $X$:
\begin{equation}
    \nabla \phi(x) \approx \nabla \phi(X) + \nabla^\mathsf{T}\nabla \phi(X) \cdot \xi(t),
\end{equation}
where $\xi = x - X$ is assumed small, and $\nabla^\mathsf{T}\nabla \phi(X)$ is the Hessian of $\phi$ near $X$. Over short times, $m\ddot \xi \approx -f(t)\nabla \phi(X)$, which can be integrated to yield $\xi(t) = \frac{1}{m} \nabla \phi(X) \int_0^t\int_0^sf(s')ds'ds$. The average force may then be expressed as
\begin{equation}
    \overline{f(t)}\nabla \phi(X) + \frac{1}{m}\overline{f(t)\left(\int_0^t \int_0^s f(s')ds' ds\right)} \nabla^\mathsf{T}\nabla \phi(X) \cdot \nabla \phi(X),
\end{equation}
where the first term vanishes due to $\overline{f} = 0$ and the second term can be written as $\frac{1}{2}K \nabla |\nabla \phi|^2$
with $K = \frac{1}{m}\overline{f(t)\left(\int_0^t \int_0^s f(s')ds' ds\right)}$. This derivation closely resembles that of Kapitza's treatment of the inverted pendulum~\cite{kapitza_dynamic_1951}, for which they considered $f(t) = F_0\cos(\omega t)$ in which case $K = \frac{F_0^2}{m\omega^2}$.

\pagebreak

\subsection{Additional Figures}

\begin{figure}[h!]
    \centering
    \includegraphics[width=\linewidth]{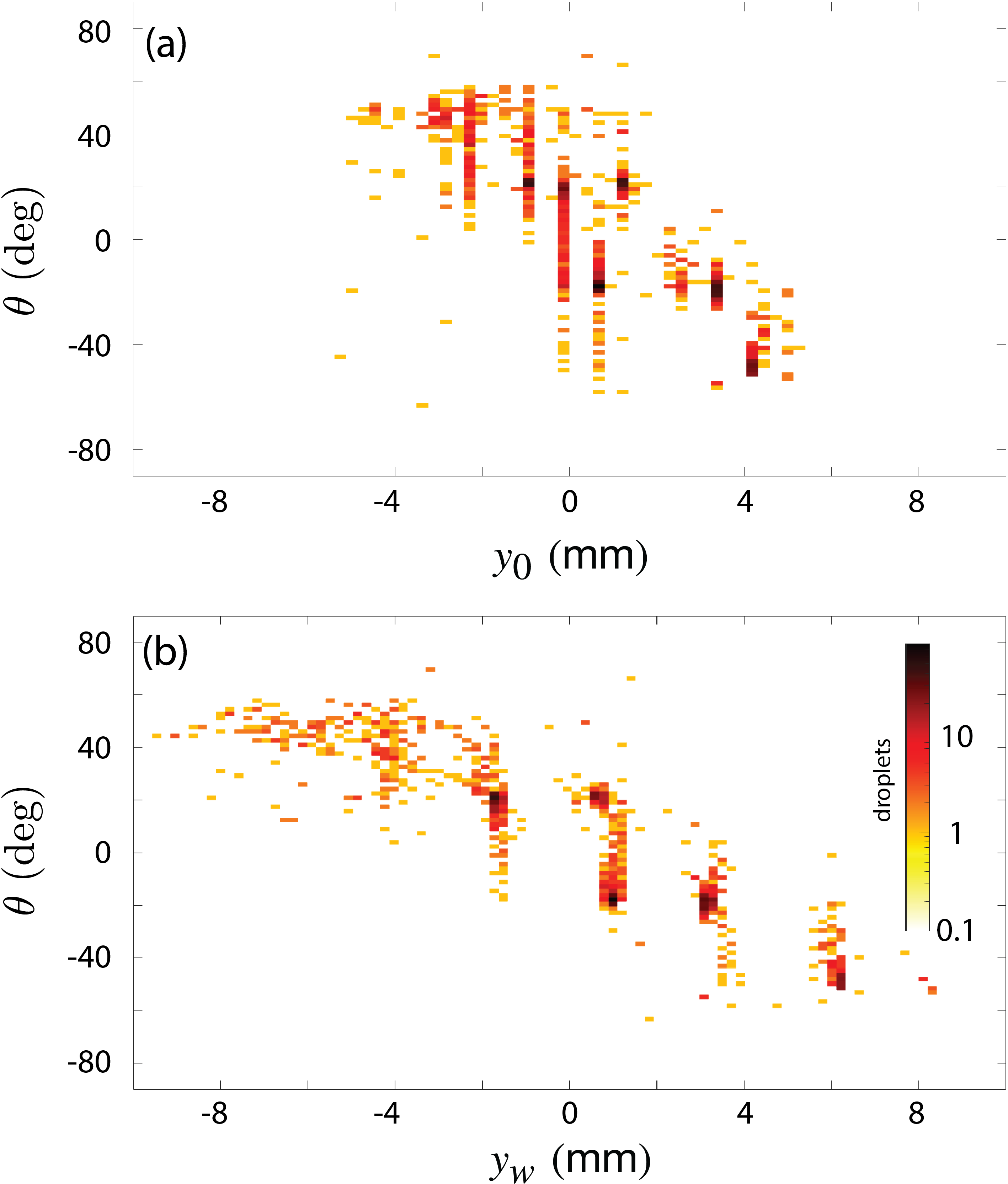}
    \caption{Experimental distribution of (a)~the impact parameter $y_0$ and (b)~the crossing distance of the drop $y_w$ (see Fig.~\ref{fig:schem}), and their associated deflection angles $\theta$. }
    \label{fig:impact_par}
\end{figure}

\begin{figure}[h!]
    \centering
    \includegraphics[width=\linewidth]{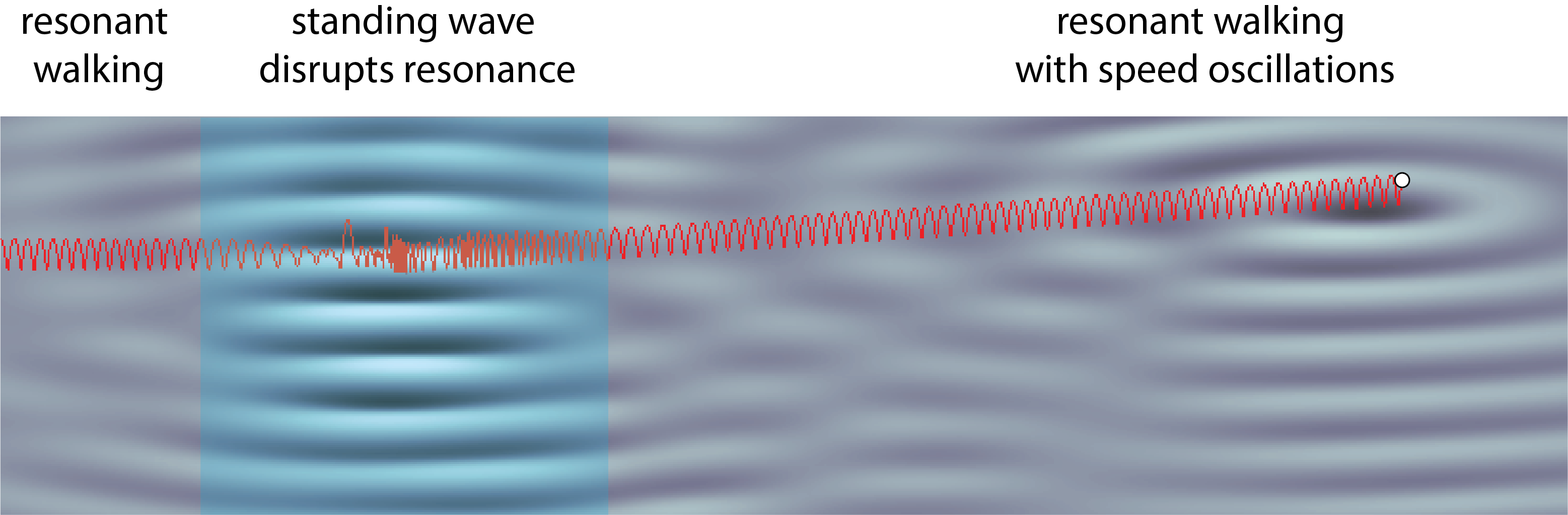}
    \caption{Simulated trajectory of a walker captures the disrupted periodicity of the droplet's bouncing above the standing Faraday wave, reproducing the experimental drop behaviour reported in Fig.~\ref{fig:highspeed}.}
    \label{fig:highspeed_sim}
\end{figure}

\end{document}